\documentstyle[12pt]{article}

\begin{document}
%\doublespace
\begin{titlepage}

\centerline{\bf STATIC BONDI ENERGY IN THE} 
\centerline{\bf TELEPARALLEL EQUIVALENT  OF GENERAL RELATIVITY}
\vskip 1.0cm
\bigskip
\centerline{\it J. W. Maluf$\,^{*}$, and J. F. da Rocha-Neto}
\centerline{\it Departamento de F\'isica}
\centerline{\it Universidade de Bras\'ilia}
\centerline{\it C.P. 04385}
\centerline{\it 70.919-970  Bras\'ilia, DF}  
\centerline{\it Brazil}
\date{}
\begin{abstract}
We consider Bondi's radiating metric in the context of the teleparallel
equivalent of general relativity (TEGR). This metric describes 
the asymptotic form of a radiating solution of Einstein's
equations. The total gravitational energy for this solution can be 
calculated by means of pseudo-tensors in the static case. 
In the nonstatic case, Bondi defines the {\it mass aspect} $m(u)$,
which describes the mass of an isolated system. In this paper 
we express Bondi's solution in asymptotically spherical 3+1 
coordinates, not in radiation coordinates, and obtain Bondi's energy 
in the static limit by 
means of the expression for the gravitational energy in the framework of 
the TEGR. We can either obtain the total energy or the energy inside 
a large (but finite) portion of a three-dimensional spacelike
hypersurface, whose boundary is far from the source.

\end{abstract}
\thispagestyle{empty}
\vfill
\par
\noindent (*) e-mail: wadih@fis.unb.br
\end{titlepage}
\newpage

\noindent {\bf I. Introduction}\par
\bigskip
The concept of energy in general relativity is considerably more 
intricate than in any other branch of physics. 
Any physical phenomena, except gravitation, is defined and described 
on a specific space-time, which is usually the flat space-time. For 
these phenomena the concept of energy can be intuitively conceived and 
mathematically realized. Generically, energy is an attribute of some 
physical system whose dynamics takes place on the space-time.
Gravitation, however, acquires a distinct status because the dynamics 
of the gravitational field is the dynamics of the space-time itself.
Consequently, the definition of the gravitational energy is not
straightforward.

The several attempts at defining the gravitational energy (pseudotensors,
quasi-local energy, actions and Hamiltonians with surface terms) all 
agree in predicting the {\it total} energy of asymptotically flat 
gravitational fields. Moreover, there seems to exist a predominant point
of view according to which the gravitational energy is not localizable,
i.e., that there does not exist a gravitational energy {\it density}.
These are probably the only two features shared by the various approaches,
which are mostly based on the metric tensor. However, the very concept of
a black hole lends support to the idea that gravitational energy
is localizable. There is no process by means of which the gravitational
mass inside a black hole can be made to vanish. 

A detailed analysis of the structure 
of the pseudotensors shows that the (covariant) gravitational 
energy-momentum tensor would have to be defined by means of 
the first derivative of the metric tensor. But it is well
known that it is not possible to write down a non-trivial covariant 
expression involving the first derivative of the metric, which captures
the energy content of the field. However, it is possible to write down
such covariant expressions with tetrads, and M\o ller noticed this
fact long ago\cite{Moller1,Moller2,Moller3}.

The question of localizability of the gravitational energy 
can be discussed in the framework of the  
teleparallel equivalent of general relativity 
(TEGR)\cite{Moller3,Hehl1,Hehl2,Hayashi}, which is an
alternative geometrical formulation of Einstein's general relativity.
The action integral of the TEGR is constructed entirely out of the
torsion tensor. The analysis of the canonical structure of the 
TEGR\cite{Maluf1} indicates the existence of a 
perfectly well defined gravitational energy density. Such existence
is possible in principle, because the TEGR is defined in terms of
tetrad fields. The torsion
tensor allows the construction of a total divergence that transforms
as a scalar (energy) density. In the 3+1 
formulation of the TEGR the integral form of the Hamiltonian 
constraint equation $C=0$ can be written as an energy equation of 
the type\cite{Maluf2}

$$C\;=\;H\,-\,E\;=\;0\;,$$

\noindent where $E$ is the gravitational energy defined by

$$E_g\;=\;{1\over {8\pi G}}\int_V d^3x\,\partial_i(eT^i)\;,\eqno(1)$$

\noindent where $e=det(e_{(k)i})$, $\lbrace e_{(k)i} \rbrace$ are
triads restricted to a three-dimensional spacelike hypersurface
$\Sigma$, and $T^i$ is the trace of the torsion tensor: 
$T^i=g^{ik}T_k=g^{ik}\,e^{(l)j}\,T_{(l)jk}$, 
$T_{(l)jk}=\partial_j e_{(l)k}-\partial_k e_{(l)j}$. $V$ is an arbitrary
three-dimensional volume of integration and $G$ is the gravitational
constant. This expression is simple and powerful. 
It has been successfully applied to rotating 
black holes\cite{Maluf3}, de Sitter space\cite{Maluf4}
and conical space-times\cite{Maluf5}. The definition of gravitational
energy in the TEGR may not be intuitively clear, but it is supported 
by its mathematical simplicity and by the applications to the
space-times listed above. The use of (1) requires only the construction
of the triads $\lbrace e_{(k)i}\rbrace$ with the appropriate boundary
conditions, and which transform under the {\it global} SO(3) group.

The torsion tensor that appears in the Hamiltonian formulation of 
the TEGR is related to the antisymmetric component of the 
connection $\Gamma^i_{jk}=e^{(m)i}\partial_j e_{(m)k}$, whose curvature
tensor is identically vanishing. Such connection  defines a space
with teleparalellism, or absolute paralelism, or else
{\it fernparallelismus}, according to Schouten\cite{Schouten}.

In this paper we investigate the energy of asymptotically flat
gravitational waves, described by Bondi's radiating metric\cite{Bondi}. 
Since the metric describes an isolated system
the application of (1) is possible as it stands, provided
we consider the metric in the 3+1 spherical coordinates 
$(t, r, \theta, \phi)$ at spacelike infinity, for which 
$t=constant$ defines a spacelike hypersurface. We note that the 
use of cartesian (rectangular) coordinates in the asymptotic limit
is necessary for the evaluation of pseudotensors out of this metric. 

We recall that the Arnowitt-Deser-Misner (ADM) energy\cite{ADM} is not
suitable for the analysis of gravitational radiation because it
gives the {\it total} energy of the space-time, 
both from the source and from the emitted
radiation, whereas the Bondi energy evaluated at null infinity 
furnishes only the energy of the source, from which it is possible to
derive the well known formula for the loss of mass.

The relevance of the definition (1) resides precisely in the fact
that we can evaluate it on a large but {\it finite} volume $V$ of the
three-dimensional spacelike hypersurface, thereby not including 
the emitted radiation outside $V$. In view of the field equations
(which are not considered here),
the energy inside $V$ turns out to be a decreasing function of time.

It is important to remark at this point that Bondi energy has been
calculated in several geometrical frameworks, by different 
aproaches\cite{Moller4,Hayward1,Hayward2,Hecht,Brown}. A
commom feature of these approaches is that they yield the {\it total}
energy of the field. In contrast, we will consider 
finite volumes of spacelike surfaces and obtain the
energy contained within large spherical surfaces of radius $r_o$
up to the ${1\over {r_o}}$ term.

In the next section we briefly describe the Lagrangian and Hamiltonian
formulations of the TEGR. In section III we compare our energy 
expression with M\o ller's expression. We show that both expressions
agree for the {\it total} gravitational energy, but 
inspite of similarities they disagree when
applied to finite volumes of the three-dimensional space.
In section IV we write Bondi's metric 
in $(t,r,\theta,\phi)$  coordinates at
infinity and proceed to carry out the construction of triads for 
the spacelike hypersurfaces $\Sigma$. There exists an infinit number of 
triads that lead to the metric restricted to the three-dimensional 
hypersurface. However, only two of them 
will be considered in detail. In
section V we calculate both the total energy of the field and the
energy contained within a large sphere of radius $r_o$. The total
energy obtained  by means of (1), in which case the integration 
is made over the whole $\Sigma$, agrees with the known 
result for the Bondi energy in the {\it static} case. We also obtain
the expression for the 
energy contained within a surface of constant radius $r_o$ 
in the asymptotic region where the metric coefficients may
be determined. \par
\bigskip
\noindent Notation: spacetime indices $\mu, \nu, ...$ and local Lorentz 
indices $a, b, ...$ run from 0 to 3. In the 3+1 decomposition latin 
indices from the middle of the alphabet indicate space indices according 
to $\mu=0,i,\;\;a=(0),(i)$. The tetrad field $e^a\,_\mu$ and
the spin connection $\omega_{\mu ab}$ yield the usual definitions
of the torsion and curvature tensors:  $R^a\,_{b \mu \nu}=
\partial_\mu \omega_\nu\,^a\,_b +
\omega_\mu\,^a\,_c\,\omega_\nu\,^c\,_b\,-\,...$,
$T^a\,_{\mu \nu}=\partial_\mu e^a\,_\nu+
\omega_\mu\,^a\,_b\,e^b\,_\nu\,-\,...$. The flat spacetime metric 
is fixed by $\eta_{(0)(0)}=-1$.               \\
\bigskip
\bigskip
\bigskip

\noindent {\bf II. The TEGR}\par
\bigskip

The Lagrangian density of the TEGR in empty spacetime is given by 

$$L(e,\omega,\lambda)\;=\;-ke({1\over 4}T^{abc}T_{abc}\,+\,
{1\over 2}T^{abc}T_{bac}\,-\,T^aT_a)\;+\;
e\lambda^{ab\mu\nu}R_{ab\mu\nu}(\omega)\;.\eqno(2)$$

\noindent where $k={1\over {16\pi G}}$, $G$ is the gravitational 
constant; $e\,=\,det(e^a\,_\mu)$, $\lambda^{ab\mu\nu}$ are 
Lagrange multipliers and $T_a$ is the trace of the torsion tensor
defined by $T_a=T^b\,_{ba}$.   The tetrad field $e_{a\mu}$ and the
spin connection $\omega_{\mu ab}$ are completely independent field
variables.  The latter is enforced to satisfy the condition of
zero curvature. Therefore this Lagrangian formulation is in no way
similar to the usual Palatini formulation, in which the spin
connection is related to the tetrad field via field equations.
Later on we will introduce the  tensor $\Sigma_{abc}$ defined by

$${1\over 4}T^{abc}T_{abc} + {1\over 2}T^{abc}T_{bac}-T^aT_a\;\equiv\;
T^{abc}\Sigma_{abc}\;.$$

The equivalence of the TEGR with Einstein's general relativity is         
based on the identity

$$eR(e,\omega)\;=\;eR(e)\,+\,
e({1\over 4}T^{abc}T_{abc}\,+\,{1\over 2}T^{abc}T_{bac}\,-\,T^aT_a)\,-\,
2\partial_\mu(eT^{\mu})\;,\eqno(3)$$

\noindent which is obtained by just substituting the arbitrary
spin connection $\omega_{\mu ab}\,=\,^o\omega_{\mu ab}(e)\,+\,
K_{\mu ab}$ in the scalar curvature tensor $R(e,\omega)$ in the
left hand side; $^o\omega_{\mu ab}(e)$ is the Levi-Civita 
connection and $K_{\mu ab}\,=\,
{1\over 2}e_a\,^\lambda e_b\,^\nu(T_{\lambda \mu \nu}+
T_{\nu \lambda \mu}-T_{\mu \nu \lambda})$ is the contorsion tensor.
The vanishing of $R^a\,_{b\mu\nu}(\omega)$, which is one of the
field equations derived from (2), implies the equivalence of 
the scalar curvature $R(e)$, constructed out of $e^a\,_\mu$ only, 
and the quadratic combination of the torsion tensor. It also
ensures that the field equation arising from the variation of
$L$ with respect to $e^a\,_\mu$ is strictly equivalent to
Einstein's equations in tetrad form. Let 
${{\delta L}\over {\delta {e^{a\mu}}}}=0$ denote the field equations
satisfied by $e^{a\mu}$. It can be shown by explicit calculations
that

$${{\delta L}\over {\delta {e^{a\mu}}}}\;=\;
{1\over 2}e \lbrace R_{a\mu}(e)\,
-\,{1\over 2}e_{a\mu}R(e)\rbrace\;.\eqno(4)$$

\noindent We refer the reader to
refs.\cite{Maluf1,Maluf2} for additional details.

Throughout this section we will be interested in asymptoticaly 
flat spacetimes. The Hamiltonian formulation of the TEGR can be 
successfully implemented if we fix the gauge $\omega_{0ab}=0$ from 
the outset, since in this case the constraints (to be 
shown below) constitute a {\it first class} set\cite{Maluf1}.
The condition $\omega_{0ab}=0$ is achieved by breaking the local
Lorentz symmetry of (2). We still make use of the residual time
dependent gauge symmetry to fix the usual time gauge condition
$e_{(k)}\,^0\,=\,e_{(0)i}\,=\,0$. Because of $\omega_{0ab}=0$,
$H$ does not depend on $P^{kab}$, the momentum canonically 
conjugated to $\omega_{kab}$. Therefore arbitrary variations of
$L=p\dot q -H$ with respect to $P^{kab}$ yields 
$\dot \omega_{kab}=0$. Thus in view of $\omega_{0ab}=0$, 
$\omega_{kab}$ drops out from our considerations. The above 
gauge fixing can be understood as the fixation of a
reference frame.

As a consequence of the above gauge fixing the canonical action 
integral obtained from (2) becomes\cite{Maluf2}

$$A_{TL}\;=\;\int d^4x\lbrace \Pi^{(j)k}\dot e_{(j)k}\,-\,H\rbrace\;,
\eqno(5)$$

$$H\;=\;NC\,+\,N^iC_i\,+\,\Sigma_{mn}\Pi^{mn}\,+\,
{1\over {8\pi G}}\partial_k (NeT^k)\,+\,
\partial_k (\Pi^{jk}N_j)\;.\eqno(6)$$

\noindent $N$ and $N^i$ are the lapse and shift functions, and 
$\Sigma_{mn}=-\Sigma_{nm}$ are Lagrange multipliers. The constraints
are defined by 

$$ C\;=\;\partial_j(2keT^j)\,-\,ke\Sigma^{kij}T_{kij}\,-\,
{1\over {4ke}}(\Pi^{ij}\Pi_{ji}-{1\over 2}\Pi^2)\;,\eqno(7a)$$

$$C_k\;=\;-e_{(j)k}\partial_i\Pi^{(j)i}\,-\,
\Pi^{(j)i}T_{(j)ik}\;,\eqno(7b)$$

\noindent with $e=det(e_{(j)k})$ and $T^i\,=\,g^{ik}e^{(j)l}T_{(j)lk}$. 
We remark that (5) and (6) are invariant under global SO(3) and
general coordinate transformations.  

If we assume the asymptotic behaviour 

$$e_{(j)k}\approx \eta_{jk}+
{1\over 2}h_{jk}({1\over r})\eqno(8)$$ 

\noindent for $r \rightarrow \infty $, then in view of 
the relation

$${1\over {8\pi G}}\int d^3x\partial_j(eT^j)\;=\;
{1\over {16\pi G}}\int_S dS_k(\partial_ih_{ik}-\partial_kh_{ii})
\; \equiv \; E_{ADM}\;\eqno(9)$$

\noindent where the surface integral is evaluated for 
$r \rightarrow \infty$, the integral form of 
the Hamiltonian constraint $C=0$ may be rewritten as

$$\int d^3x\biggl\{ ke\Sigma^{kij}T_{kij}+
{1\over {4ke}}(\Pi^{ij}\Pi_{ji}-{1\over 2}\Pi^2)\biggr\}
\;=\;E_{ADM}\;.\eqno(10)$$

\noindent The integration is over the whole three dimensional
space. Given that $\partial_j(eT^j)$ is a scalar  density,
from (9) and (10) we define the gravitational
energy density enclosed by a volume V of the space as

$$E\;=\;{1\over {8\pi G}}\int_V d^3x\partial_j(eT^j)\;.\eqno(1)$$  

\noindent It must be noted that $E$ depends only on the triads
$e_{(k)i}$ restricted to a three-dimensional spacelike hypersurface;
the inverse quantities $e^{(k)i}$ can be written in terms of 
$e_{(k)i}$. From the identity (4) we observe that the dynamics of
the triads does not depend on $\omega_{\mu ab}$. Therefore $E_g$
given above does not depend on the fixation of any gauge for
$\omega_{\mu ab}$. We briefly remark that the
reference space which defines the zero of energy has been 
discussed in ref.\cite{Maluf3}. 

We make now the important assumption that general form of the canonical
structure of the TEGR is the same for any class of space-times, 
irrespective of the peculiarities of the latter (for the de Sitter
space\cite{Maluf4}, for example, there is an {\it additional}
term in the Hamiltonian constraint $C$).
Therefore we assert that the integral form of the Hamiltonian 
constraint equation can be written as $C=H-E=0$ for {\it any} 
space-time, and that (1) represents the gravitational 
energy for arbitrary space-times with any topology.                       

Before closing this section, let us recall that M\"uller-Hoissen and
Nitsch\cite{MHN} and $\;\;\;$
Kopczy\'nski\cite{Kop} have shown that in general
the theory defined by (2) faces difficulties with respect to the 
Cauchy problem. They have shown that in general six components of
the torsion tensor are not determined from the evolution of the 
initial data. On the other hand, the constraints of the  theory
constitute a first class set provided we fix the six quantities
$\omega_{0ab}=0$ {\it before varying the action}\cite{Maluf1}. 
This condition 
is mandatory and does not merely represent one particular gauge
fixing of the theory. Since the fixing of $\omega_{0ab}$ yields a
well defined theory with first class constraints, we cannot 
assert that the field configurations of the
latter are gauge equivalent to configurations
whose time evolution is not precisely determined. The requirement of
local SO(3,1) symmetry plus the addition of 
$\lambda^{ab\mu \nu}R_{ab\mu\nu}(\omega)$ in (2) has the ultimate
effect of discarding the connection.
Although we have no proof, we believe that
the two properties above (the failure of the Cauchy problem and the
fixation of $\omega_{0ab}=0$) are related to each other.      

Constant rotations constitute a  basic feature of the teleparallel   
geometry. According to M\o ller\cite{Moller2}, in the framework
of the abolute parallelism tetrad fields, together with the
boundary conditions, uniquely determine a {\it tetrad lattice}, 
apart from an arbitrary {\it constant rotation of the tetrads in the 
lattice}.  \\
\bigskip
\bigskip

\noindent {\bf III. M\o ller's energy expression}\par
\bigskip
M\o ller carried out several investigations regarding the localizability
of the gravitational energy. He faced difficulties in establishing
a covariant expression using the metric tensor\cite{Moller2}, 
and because of this he arrived at an expression through the use of 
tetrads\cite{Moller2,Moller3}. According to M\o ller, this latter
expression still has a difficulty in that it is not invariant under
{\it local} Lorentz transformations. 
It is very instructive to compare expression (1) with M\o ller's
expression. For the sake of this comparison,
we will put aside the difficulty regarding the noninvariance with
respect to local Lorentz transformations. 

M\o ller presents an expression for the energy-momentum of the
gravitational field. However we will only consider the energy
expression. Translating into our notation, M\o ller's energy reads

$$E\;=\;-\int d^3x\,
\partial_\lambda U_0\,^{0\lambda}\;,\eqno(11)$$

\noindent where the potential in the integrand is given by

$$U_\mu\,^{\nu\lambda}\;=\;
{1\over {8\pi G}}e\,[e^{a\nu}\nabla_\mu e_a\,^\lambda+
(\delta^\nu_\mu e^{a\lambda}-\delta^\lambda_\mu e^{a\nu})
\nabla_\sigma e_a\,^\sigma]\eqno(12)\;.$$

\noindent In contrast with the notation of the previous section,
all geometrical quantities in equations (11-15) are 
{\it four-dimensional} quantities. In (12) $\;\nabla\;$ represents 
the covariant derivative with respect to the Christoffel symbols
$\Gamma^\lambda_{\mu\nu}$.

M\o ller's energy can be first rewritten as

$$E\;=\;{1\over {8\pi G}}
\int d^3x\,\partial_i(e\,e^{ak}\nabla_ke_a\,^i)\;.\eqno(13)$$

\noindent By means of the identity

$$\nabla_k e_{aj}=\partial_k e_{aj}-\Gamma^\sigma_{kj}e_{a\sigma}\equiv
-^o\omega_k\,^b\,_a e_{bj}\;,\eqno(14)$$

\noindent where $^o\omega_{\mu ab}$ is the Levi-Civita connection,

$$^o\omega_{\mu ab}=-{1\over 2}e^c\,_\mu(\Omega_{abc}-\Omega_{bac}-
\Omega_{cab})\;,$$

$$\Omega_{abc}=e_{a\nu}(e_b\,^\mu \partial_\mu e_c\,^\nu-
e_c\,^\mu \partial_\mu e_b\,^\nu)\;,$$

\noindent we can further rewrite expression (13) as

$$E\;=\;{1\over {8\pi G}}\int d^3x\,\partial_i(ee^{ai}e^{bj}\,
^o\omega_{jab})\;.\eqno(15)$$

Up to this point $\{e_{a\mu}\}$ are tetrads of the four-dimensional
space-time. In order to compare (15) with (1) let us impose
the time gauge $e^{(0)}\,_k=e_{(j)}\,^0=0$ and establish the 3+1
decomposition of the tetrads as in \cite{Maluf1,Maluf6}. Then the
integrand on the right hand side of (15) can be rewritten as

$$^4e\,^4e^{ai}\,^4e^{bj}\,^o\omega_{jab}(^4e)=
Nee^{(m)i}e^{(n)j}\,^o\omega_{j(m)(n)}(e)-
e(e^{ai}N^j-e^{aj}N^i)\,^o\omega_{j(0)(m)}$$

\noindent $N$ and $N^i$ are the lapse and shift functions and 
$\{^4e^{a\mu}\}$ are tetrads of the four-dimensional space-time. In
terms of triads restricted to a three-dimensional spacelike surface we
have

$$E\;=\;{1\over {8\pi G}}\int d^3x 
\partial_i[Nee^{(m)i}e^{(n)j}\,^o\omega_{j(m)(n)}-
e(e^{(m)i}N^j-e^{(m)j}N^i)^o\omega_{j(0)(m)}]\;.\eqno(16a)$$

Comparison with (1) can now be made if we make use of the
{\it identity}\cite{Maluf7}

$$\partial_i(e e^{(m)i} e^{(n)j}\,^o\omega_{j(m)(n)})\equiv
\partial_i(eT^i)\;,$$

\noindent where the right hand side above is the same as in (1). 
M\o ller energy can be finally written as

$$E\;=\;{1\over {8\pi G}}\int d^3x \partial_i[NeT^i-
e(e^{(m)i}N^j-e^{(m)j}N^i)\,^o\omega_{j(0)(m)})]\;.\eqno(16b)$$

\noindent Recall that we are ignoring {\it local} Lorentz transformations.

Besides the appearance of extra terms involving $^o\omega_{j(0)(m)}$ on
the right hand side of (16a,b), there is also the crucial presence of the
lapse function $N$ multiplying $eT^i$. Therefore even for configurations
of the gravitational field for which the second term on the right 
hand side of (16) does not contribute (if, say, $N^i=0$, as for the
Schwarzschild solution) expressions (1) and (16) will lead to
different results when applied to finite volumes of the three-dimensional
space. Moreover, because of the presence of the lapse function, 
(16) is not invariant under time reparametrizations:
$N'({x'}^0)={{\partial {x'}^0} \over {\partial x^0}}N(x^0)$. Thus for
a finite volume of integration (16b) does not remain invariant
under this reparametrization.

In the Einstein-Cartan theory the connection $^o\omega_{j(0)(m)}$ 
can be expressed in terms of the momenta canonically conjugated to
$e_{(m)i}$. In the notation of \cite{Maluf6} it is given by
$^o\omega_{j(0)(m)}={1\over {2e}}(\pi_{(m)j}-{1\over 2}e_{(m)j}\pi)$
(see equation 12 of \cite{Maluf6}). In this context (16b) can
be rewritten as

$$E\;=\;{1\over {8\pi G}}\int d^3x\partial_i \lbrack NeT^i-
{1\over 2} e^{(m)i}N^j \pi_{(m)j}\rbrack\;.$$

\noindent The expression above is exactly the energy expression for
the Einstein-Cartan theory (see equation 21 of \cite{Maluf6}), assuming
that the gravitational energy is obtained from integration of surface
terms of the total Hamiltonian. This expression is also very similar
to (i) the integral of the surface terms in equation 6, and (ii)
the energy expression considered by Nester\cite{Nester} in the 
analysis of the positivity of the gravitational energy
(equation (3.15) of \cite{Nester}).
All definitions of gravitational energy
considered above agree for the total gravitational energy.    \\
\bigskip
\bigskip
\bigskip

\noindent {\bf IV. Bondi's radiating metric and the associated triads}\par
\bigskip
Bondi's metric is a not an exact solution of Einstein's equations.
It describes the asymptotic form of a radiating solution.
In terms of radiation coordinates $(u,r,\theta,\phi)$, where $u$
is the retarded time and $r$ is the luminosity distance, 
Bondi's radiating metric is written as

$$ds^2\;=\;-\biggl( {V\over r} e^{2\beta}- U^2\,r^2 e^{2\gamma}\biggr)du^2
-2e^{2\beta}du\,dr - 2U\,r^2\,e^{2\gamma}du\,d\theta$$

$$+r^2 \biggl( e^{2\gamma}\,d\theta^2 + 
e^{-2\gamma}\,sin^2\theta\,d\phi^2\biggr)\;.\eqno(17)$$

\noindent The metric above is such that the surfaces
for which $u=constant$ are null hypersurfaces. Each null radial (light)
ray is labelled by particular values of $u, \theta$ and $\phi$. At
spacelike infinity $u$ takes the standard form $u=t-r$. The four 
quantities appearing in (17), $V, U, \beta$ and $\gamma$ are functions of
$u, r$ and $\theta$. Thus (17) displays axial symmetry. A more general
form of this metric has been given by Sachs\cite{Sachs}, who  showed 
that the most general metric tensor describing asymptotically flat 
gravitational waves depends on six functions of the coordinates.

The functions in (17) satisfy the following asymptotic behaviour:

$$\beta\;=\;-{c^2\over {4r^2}}+...$$

$$\gamma\;=\;{ c \over r}+O({1\over {r^3}})+...$$

$${V\over r}\;=\;1\,-\,{{2M}\over r}\,
-\,{1\over r^2}\biggl[ {{\partial d}\over {\partial \theta}} +
d\,cot\theta-\biggl({{\partial c}\over{\partial \theta}}\biggr)^2
-4c\biggl({{\partial c}\over {\partial \theta}}\biggr)cot\theta-
{1\over 2}\,c^2\biggl(1+8cot^2\theta\biggr)\biggr]+...$$

$$U\;=\;{1\over r^2}\biggl( {{\partial c}\over {\partial \theta}}
+2c\,cot\theta\biggr)+ 
{1\over{r^3}}\biggl(2d+3c\,{{\partial c}\over{\partial \theta}}
+4c^2\,cot\theta\biggr)+...$$

\noindent where $M=M(u,\theta)$ and $d=d(u,\theta)$ are the mass aspect
and the dipole aspect, respectively, and from the function $c(u,\theta)$
we define the news function ${{\partial c(u,\theta)} \over {\partial u}}$.

The application of (1) to Bondi's metric requires transforming it to
coordinates $t, r, \theta$ and $\phi$ for which $t=constant$ defines a
space-like hypersurface. Before proceeding, we recall that the 
analysis of (17) in $t,x,y,z$ coordinates has already been performed by 
Goldberg\cite{Goldberg}, in the investigation of the asymptotic invariants
of gravitational radiation fields. Therefore we carry out a coordinate
transformation such that the new timelike coordinate is given by
$t=u+r$. We arrive at

$$ds^2\;=\; -\biggl({V\over r}e^{2\beta}-U^2\,r^2\,e^{2\gamma}\biggr)dt^2
-2U\,r^2\,e^{2\gamma}dt\,d\theta
+2\biggl[e^{2\beta}\biggl({V\over r}-1\biggr)-
U^2\,r^2\,e^{2\gamma}\biggr]dr\,dt$$

$$+\biggl[e^{2\beta}\biggl(2-{V\over r}\biggr)+
U^2\,r^2\,e^{2\gamma}\biggr]dr^2
+2U\,r^2\,e^{2\gamma}dr\,d\theta + 
r^2\biggl( e^{2\gamma}d\theta^2+
e^{-2\gamma}\,sin^2\theta\,d\phi^2\biggr)\;.\eqno(18)$$

\noindent Therefore the metric restricted to a three-dimensional
spacelike hypersurface is given by

$$ds^2\;=\;\biggl[ e^{2\beta}\biggl(2-{V\over r}\biggr)+
U^2\,r^2\,e^{2\gamma}\biggr] dr^2+
2U\,r^2\,e^{2\gamma}\,dr\,d\theta$$

$$+r^2\biggl(e^{2\gamma}\,d\theta^2+
e^{-2\gamma}\,sin^2\theta\,d\phi^2\biggr)\;.\eqno(19)$$

We must consider triads that  correspond to the metric above.
The construction of triads, in general, is a nontrivial step. If
in a given coordinate system the metric tensor is diagonal, then
the construction of triads is a relatively simple procedure.
One must only make sure that the triads satisfy the appropriate
boundary conditions at infinity. Recall that in order to obtain
expression (9) for the ADM energy the triads must have the
appropriate asymptotic behaviour given by equation (8).

The metric tensor (19) has an off-diagonal element, and this
fact adds a bit of complication in the construction of triads. 
Nevertheless we can immediately write down two sets of triads that
lead to this metric. They are given by

$$e_{(k)i}\;=\;\pmatrix{ A\,sin\theta\,cos\phi+B\,cos\theta\,cos\phi&
rC\,cos\theta\,cos\phi & -rD\,sin\theta\,sin\phi\cr
A\,sin\theta\,sin\phi+B\,cos\theta\,sin\phi &
rC\,cos\theta\,sin\phi & rD\,sin\theta\,cos\phi\cr
A\,cos\theta - B\,sin\theta & -rC\,sin\theta & 0\cr}\;,\eqno(20)$$

\noindent where

$$A\;=\;e^\beta \sqrt{2-{V\over r}}\;,\eqno(21a)$$

$$B\;=\;r\,U\,e^\gamma\;,\eqno(21b)$$

$$C\;=\;e^\gamma\;,\eqno(21c)$$

$$D\;=\;e^{-\gamma}\;,\eqno(21d)$$

\noindent and

$$e_{(k)i}\;=\;\pmatrix{ A'\,sin\theta\,cos\phi&
rB'\,cos\theta\,cos\phi+rC'\,sin\theta\,cos\phi&-rD'\,sin\theta\,sin\phi\cr
A'\,sin\theta\,sin\phi&rB'\,cos\theta\,sin\phi+rC'\,sin\theta\,sin\phi&
rD'\,sin\theta\,cos\phi\cr
A'\,cos\theta&-rB'\,sin\theta + rC'\,cos\theta&0\cr}\;,\eqno(22)$$

\noindent where

$$A'\;=\;\bigg[ e^{2\beta}\biggl( 2-{V\over r}\biggr)+
U^2\,r^2\,e^{2\gamma}\biggr]^{1\over 2}\;,\eqno(23a)$$

$$B'\;=\;{1\over A'}\,e^{\beta + \gamma}\sqrt{2-{V\over r}}\;,\eqno(23b)$$

$$C'\;=\;{1\over A'}\, U\,r\,e^{2\gamma}\;,\eqno(23c)$$

$$D'\;=\;e^{-\gamma}\;.\eqno(23d)$$

\noindent It is easy to see that both (20) and (22) yield the 
metric tensor (19) through the relation  $e_{(i)j}e_{(i)k}=g_{jk}$.
They are related by a {\it local} SO(3) transformation. 

Triads given by (20) and (22)
are the {\it simplest} sets of triads that satisfy the two basic
requirements: ({\bf i}) the triads must have the asymptotic
behaviour given by (8); ({\bf ii}) by making the physical parameters
of the metric vanish we must have $T_{(k)ij}=0$ everywhere.
In the present case if we make $M=d=c=0$ both (20) and (22) 
acquire the form

$$e_{(k)i}\;=\;\pmatrix{sin\theta\,cos\phi &
r\,cos\theta\,cos\phi &
-r\,sin\theta\,sin\phi\cr
sin\theta\,sin\phi &
r\,cos\theta\,sin\phi &
r\,sin\theta\,cos\phi\cr
cos\theta &
-r\,sin\theta & 0\cr}\;.\eqno(24)$$

\noindent In cartesian coordinates the expression above can be 
reduced to the diagonal form $\;\;$
$e_{(k)i}(x,y,z)=\delta_{ik}$. The requirement (ii) above is 
essentialy equivalent to the establishment of a reference 
space, as discussed
in \cite{Maluf3}. Without the notion of a reference space we
cannot define gravitational energy. Note that by a suitable choice
of a local SO(3) rotation we can make the flat space triad (24) 
satisfy the requirement (i), but not (ii).

It is possible to show that (20) and (22) are particular cases
of an infinit set of triads that satisfy both requirements above. 
It is given by

$$e_{(k)i}=\pmatrix{ 
{\cal A}\,sin\theta cos\phi+{\cal B}\,cos\theta cos\phi&
r{\cal C}\,cos\theta cos\phi+r{\cal D}\,sin\theta cos\phi&
-re^{-\gamma} \,sin\theta sin\phi \cr
{\cal A}\,sin\theta sin\phi+{\cal B}\,cos\theta sin\phi&
r{\cal C}\,cos\theta sin\phi+r{\cal D}\,sin\theta sin\phi&
re^{-\gamma} \,sin\theta cos\phi\cr
{\cal A}\,cos\theta -{\cal B}\,sin\theta&
-r{\cal C}\,sin\theta + r{\cal D}\,cos\theta & 0\cr}\;,\eqno(25)$$

\noindent with the following definitions:

$${\cal A}\;=\;\sqrt{
e^{2\beta}\biggl(2-{V\over r}\biggr)+
U^2r^2\biggl(e^{2\gamma}-{\bf b}^2\biggr)}\;,$$

$${\cal B}\;=\;{\bf b}\,Ur\;,$$

$${\cal C}\;=\;\sqrt{e^{2\gamma}-{\bf d}^2\,U^2r^2 }\;,$$

$${\cal D}\;=\;{\bf d}\,Ur\;,$$

\noindent where ${\bf b}$ and ${\bf d}$ are arbitrary, 
dimensionless functions that must satisfy

$${\bf b}\,=\,\sqrt{e^{2\gamma} -{\bf d}^2\,U^2r^2}+
{\bf d}\,e^{-\gamma+\beta}\sqrt{2-{V\over r}}$$

\noindent By making ${\bf d}=0$ we obtain (20), and 
${\bf b}=0$ leads to (22).\par
\bigskip

From the point of view of the TEGR triads given by (20) and (22)
are physically inequivalent (that is, they are not
gauge equivalent), because we have
seen that the Hamiltonian formulation established by eq.(5) 
is not invariant under the local SO(3) group, but rather under 
the global SO(3). In the TEGR the torsion tensor describes the way
in which the space-time is deformed. The latter is thus considered
as a continuum with microstructure\cite{Hehl2}.
Therefore the same space,
defined uniquely by the metric tensor, may be deformed in several
ways, according to the manner one defines the triads. This 
is essentially the geometrical meaning of the noninvariance of the
TEGR under local SO(3) transformations.

In the Hamiltonian formulation of the TEGR the basic geometrical field 
variable is the triad, not the metric tensor. Any set of 
triads should be ruled out on physical grounds, i.e., if they lead to 
incorrect physical statements concerning the energy content of 
the gravitational field.   

In the next section we will obtain
the expressions for the gravitational energy arising from (20) and 
(22). These expressions are quite different.
Although the expression corresponding to (20) is simpler, as we
will see, we have no definite experimental evidence in favour of it.

\bigskip 
\bigskip
\noindent {\bf V. Gravitational radiation energy}\par
\bigskip
In this section we will apply expression (1) both to (20) and (22).
Since the two triads display distinct geometrical properties, we 
expect to obtain different expressions for the energy density
${1\over {8\pi}}\partial_i(eT^i)$ (we will make the gravitational
constant $G=1$). Our analysis is meaningful
only in the asymptotic region of large values of the radial distance. 
However, we have no reason to expect (20) and (22) to yield
different expressions for the {\it total} energy of the field. In 
fact, as we will see, they yield the same (expected) expression.

As we mentioned earlier, the significance
of the present approach to the analysis of gravitational
radiation fields resides in the fact that we can evaluate the 
gravitational energy inside a large but finite portion of a 
three-dimensional spacelike surface. In other words, by means of
Gauss law (1) can be evaluated over a surface far from the source
of radiation, in the asymptotic limit where the metric components
are precisely determined. Specifically, we will evaluate (1) inside
a large sphere of radius $r_o$. The time evolution of the metric
field will determine the time dependence of this energy, and 
consequently the energy radiated out of it. 

In order to
calculate $T^i=g^{ik}g^{mj}e_{(l)m}T_{(l)jk}$ we need the inverse
metric $g^{ij}$. In terms of the definitions (21), it is given by

$$g^{ij}\;=\;\pmatrix{ {1\over {A^2}}& -{B\over {r\,e^{\gamma}\,A^2}}&0\cr
-{B\over {r\,e^{\gamma}\,A^2}}&
{1\over {r^2\,e^{2\gamma}}} (1+{B^2\over A^2} )&0\cr
0&0&{1\over {r^2\,e^{-2\gamma}\,sin^2\theta}}\cr}\;.\eqno(26)$$

\bigskip
We will initially consider the set of triads given by (20). In the
following, a comma after a field quantity indicates a derivative:
$A,_1$  and $A,_2$ indicate derivative with of $A$ respect to $r$ and  
$\theta$, respectively. The torsion components for (20) are given by

$$T_{(1)12}=cos\theta\,cos\phi\,(C+rC,_1-A-B,_2)
+sin\theta\,cos\phi\,(-A,_2+B)\;,$$

$$T_{(1)13}=sin\theta\,sin\phi\,(A-D-rD,_1)+cos\theta\,sin\phi\,B\;,$$

$$T_{(1)23}=sin\theta\,sin\phi\,(-rD,_2)+cos\theta\,sin\phi\,(rC-rD)\;,$$

$$T_{(2)12}=cos\theta\,sin\phi\,(C+rC,_1-A-B,_2)
+sin\theta\,sin\phi\,(-A,_2+B)\;,$$

$$T_{(2)13}=sin\theta\,cos\phi\,(-A+D+rD,_1)+cos\theta\,cos\phi\,(-B)\;,$$

$$T_{(2)23}=sin\theta\,cos\phi\,(rD,_2)+cos\theta\,cos\phi\,(-rC+rD)\;,$$

$$T_{(3)12}=sin\theta\,(A+B,_2-C-rC,_1)+cos\theta\,(-A,_2+B)\;,$$

$$T_{(3)13}=T_{(3)23}=0\;.$$

Since we are interested in calculating the energy inside a surface of
constant radius, only $T^1$ will be considered. By Gauss law, the
expression of this energy is given by

$$E\;=\;{1\over {8\pi}}\int_S d\theta\,d\phi\,eT^1\;,\eqno(27)$$

\noindent where $S$ is a surface of fixed radius $r_o$, assumed to 
be large as compared with the dimension of the source, and the
determinant $e$ is given by $e=r^2\,A\,sin\theta$. After a long but
otherwise straightforward calculation we arrive at

$$E_I \;=\;{ {r_o}\over 4}\int_0^\pi\,d\theta\,
\biggl\{ sin\theta\biggl[e^\gamma + e^{-\gamma}-
{2\over A}  \biggr]+
{1\over A}\, {{\partial}\over {\partial \theta}}(Ur\,sin\theta)\biggr\}
\;.\eqno(28)$$
	  
\noindent with $A$ defined by (21a). Let us note that the field 
quantities appearing in (28) are functions of $u=t-r$: 
$M=M(t-r,\theta), c=c(t-r,\theta)$, etc. Therefore, once these
functions are known, one can explicitely calculate the variation of
$E_I$ with respect to the time $t$, however only in the limit where 
the metric components are precisely determined.

Unfortunately the expansion of $E_I$ up to terms in $1\over {r_o}$, 
making use of the aymptotic behaviour of $U, V, \beta$ and $\gamma$,
yields no simple expression. It is given by

$$E_I\;=\;{1\over 2}\int_0^\pi d\theta\,sin\theta\,M
-{1\over {4r_o}}\int_0^\pi d\theta\,sin\theta
\biggl[ \biggl(  {{\partial c}\over {\partial \theta}} \biggr)^2\,
+4c\biggl({{\partial c}\over {\partial \theta}}\biggr)cot\theta
+4c^2\,cot^2\theta\biggr]$$

$$-{1\over {4r_o}}\int_0^\pi d\theta\,M\,{\partial \over {\partial \theta}}
\biggl[sin\theta\biggl( {{\partial c}\over {\partial \theta}}
+2c\,cot\theta \biggr)\biggr]\;.\eqno(29)$$

\noindent In the calculation above we have assumed
that $U(\theta)sin\theta=d(\theta)sin\theta=0$ when $\theta=0$ or
$\pi$.

We observe that $E_I$ yields Bondi energy in the limit 
$r \rightarrow \infty$ in the {\it static} case (i.e., when $M$ 
is a function of $\theta$ only), so that $E_I$ is the total
gravitational energy. However, in the nonstatic case
an expression for the loss of mass due to gravitational radiation
can be obtained from (29). This is one major result of our analysis.

Let us recall that Bondi's {\it mass aspect} $m(u)$,

$$m(u)\;
=\;{1\over 2}\int_0^\pi d\theta\,sin\theta\,M(u,\theta)\;,\eqno(30)$$

\noindent depends on the null foliation used. The mass aspect $m(u)$
can be understood as a mass associated to each null cone determined
by the equation $u=constant$.  Since the limit $r\rightarrow \infty$
corresponds to $u\rightarrow -\infty$ for finite $t$, we see once
again that in this limit $E_I$ gives the total energy because it
corresponds to the initial value of the Bondi energy.
\bigskip

We will consider next the second set of triads, eq.(22). The components
of the torsion tensor resulting from the latter are given by

$$T_{(1)12}=cos\theta\,cos\phi\,(B'+rB',_1-A')+
sin\theta\,cos\phi\,(C'+rC',_1-A',_2)\;,$$

$$T_{(1)13}=sin\theta\,sin\phi\,(A'-D'-rD',_1)\;,$$

$$T_{(1)23}=cos\theta\,sin\phi\,(rB'-rD')+
sin\theta\,sin\phi\,(rC'-rD',_2)\;,$$

$$T_{(2)12}=cos\theta\,sin\phi\,(B'+rB',_1-A')+
sin\theta\,sin\phi\,(C'+rC',_1-A',_2)\;,$$

$$T_{(2)13}=-sin\theta\,cos\phi\,(A'-D'-rD',_1)\;,$$

$$T_{(2)23}=-cos\theta\,cos\phi\,(rB'-rD')-
sin\theta\,cos\phi\,(rC'-rD',_2)\;,$$

$$T_{(3)12}=-sin\theta\,(B'+rB',_1-A')+
cos\theta\,(C'+rC',_1-A',_2)\;,$$

$$T_{(3)13}\,=\,T_{(3)23}\,=0\;.$$

\noindent As in the previous case, we are interested  in
calculating the energy in the interior of a surface of constant
radius $r_o$. Therefore only the knowledge of
$T^1$ will be necessary. After a long calculation we first arrive at

$$eT^1\;=\;-{{r\,sin\theta}\over A}\biggl\{e^{-2\gamma}\biggl[
B'{\partial \over {\partial r}}(rB')
+C'{\partial \over {\partial r}}(rC')-A'B'
-C'{{\partial A'}\over {\partial \theta}}\biggr]$$

$$+\;e^{2\gamma} \biggl[-A'e^{-\gamma}-
r{{\partial \gamma}\over \partial r}e^{-2\gamma}
+\;e^{-2\gamma}\biggr] \biggr\}$$

$$-{{rB\,sin\theta}\over A}\biggl[ C'+
{{\partial \gamma}\over {\partial \theta}}\,e^{-\gamma}\biggr]
-{{rB\,cos\theta}\over A}\biggl[ B'-e^{-\gamma}\biggr]\;,\eqno(31)$$

\noindent where the primed quantities are given by (23).
It is not straightforward to put the expression above in a 
simplified form. After some rearrangements we can finally  write
the energy expression (27) as

$$E_{II}\;=\;{r_o\over 4}\int_0^\pi d\theta\,{1\over A}\biggl\{sin\theta
\biggl[ e^\gamma A'+e^{-2\gamma}A'B'-2+
e^{-2\gamma}{{\partial A'}\over {\partial \theta}}C'-BC'
-Be^{-\gamma}  {{\partial \gamma} \over {\partial \theta}}\biggr]$$

$$-B\,cos\theta\biggl[ B'-e^{-\gamma}\biggr] \biggr\}\;.\eqno(32)$$

\noindent Like equation (28), this expression represents the energy
enclosed by a large spherical surface of radius $r_o$.
Expanding the expression above up to the first 
power of $1\over {r_o}$ we find

$$E_{II}\;=\;{1\over 2}\int_0^\pi d\theta\,M\,sin\theta-
{1\over {4r_o}}\int_0^\pi d\theta\,sin\theta \biggl[
3M^2+{5\over 2}\biggl( {{\partial c}\over {\partial \theta}}\biggr)^2\,
+10c\biggl({{\partial c}\over {\partial \theta}} \biggr)cot\theta$$

$$+8c^2cot^2\theta-\biggl({{\partial M}\over {\partial \theta}}\biggr)
\biggl({{\partial c}\over {\partial \theta}}+2c\,cot\theta\biggr)
\biggr]
-{1\over {4r_o}}\int_0^\pi d\theta\,cos\theta\biggl[2c
\biggl({{\partial c}\over {\partial \theta}}\biggr)+
4c^2\,cot\theta\biggr]\;.\eqno(33)$$

\noindent We are again assuming 
$U(\theta)sin\theta=d(\theta)sin\theta=0$ for $\theta=0,\pi$.

We observe that in the limit $r \rightarrow \infty\;\;$ $E_{II}$ also 
gives the total energy. As before, for a finite (but sufficiently
large) value of $r_o$ we can compute the loss of mass due to
gravitational radiation, once the functions $M$ and $c$ are
known in the asymptotic region.   \\

\bigskip
\bigskip
\noindent {\bf VI. The selection of triads}\par
\bigskip
In section IV we obtained an infinit set of triads that yield
the three-dimensional spacelike section of Bondi's metric, and in
the previous section we considered in detail only the simplest
constructions. Of course simplicity is a major feature of 
physical systems, but we are really in need of experimental 
evidence that leads to a definite description. 
We need actual realizations of the quantities $M(r-t,\theta)$,
$c(r-t,\theta)$, $d(r-t,\theta)$ and experimental evidence
on how the energy is radiated away 
in order to arrive simultaneously at the correct energy expression
arising from (27) and at the definite expression of $e_{(k)i}$.

However, we can envisage two possible types of 
conditions on the triads that associate a unique 
triad with the three-dimensional metric tensor.

The first condition regards the energy content of the gravitational
field. If we stick to the point of view according to which 
physical systems in nature have a tendency to be in  states of
minimum energy, then the correct triad for the spacelike section
of Bondi's metric is the one that minimizes expression (27)
for all possible constructions of $e_{(k)i}$. 
By means of this criterium we consider triads given by (25),
or any further construction that complies with the two conditions 
stated in section IV, and ask which one yields the smaller 
value of energy
contained within a surface of constant radius, in similarity with
the calculations of (29) and (33). Unfortunately this analysis
cannot be carried out unless $M$, $c$ and $d$ are known.

Certainly one can ask whether only (27) should be minimized or
the energy density should be everywhere a minimum. In the context
of Bondi's metric the latter possibility cannot be 
considered, because the
metric is valid only in the asymptotic limit, but in the general
case it is an open question that must be carefully addressed. 

The second condition takes into account equation (8):
we require the triads to have the asymptotic behaviour 
determined by (8)  with a {\it symmetric} tensor $h_{jk}=h_{kj}$.
Again, one has to find out of (25) which realization of 
$e_{(i)k}$ in cartesian coordinates complies with this criterium.
This condition may be understood as a {\it rotational
gauge condition}. Note that as it stands, $h_{jk}$ in equation (8)
is not required to be symmetric (in equation (9) only the
symmetrical part contributes). By explicit calculations we observe
that neither (20) nor (22) satisfy this second condition.

The two conditions above may not be mutually excluding. On the
contrary, they may lead to the same triad. The determination of the
correct triad is certainly an essential and crucial 
issue of the theory
and will be further investigated elsewhere in the general case,
with special attention to Bondi's metric, in the light of the
conditions above.

We observed that both (29) and (33) yield the same total energy. 
This is also the case if we carry out the calculations with
a more complicated triad, whether belonging to (25) or not,
which is related to (20) or (22) by a local SO(3) transformation
with an appropriate asymptotic behaviour. Let us consider a
local SO(3) transformation given by

$$\tilde e^{(k)}\,_i(x)\;
=\;\Lambda^{(k)}\,_{(l)}(x)\,e^{(l)}\,_i(x)\;.\eqno(34)$$

\noindent Under (34) the energy expression (1) transforms as

$$\tilde E\;=\; E\,+{1\over{8\pi}}\int_V d^3x\partial_i\lbrack
eg^{ik}\Lambda^{(l)}\,_{(m)}e^{(m)j}\,(
e_{(n)k}\partial_j\Lambda_{(l)}\,^{(n)}-
e_{(n)j}\partial_k\Lambda_{(l)}\,^{(n)})\rbrack\;.\eqno(35)$$

\noindent Expression (35) can be best analysed if we consider
an infinitesimal rotation.
We assume that in the limit $r\rightarrow \infty$
the SO(3) elements have the asymptotic behaviour

$$\Lambda^{(k)}\,_{(l)}\, \approx \;\delta^{(k)}_{(l)}\;+\;
^0\omega^{(k)}\,_{(l)}\;+\;
^1\omega^{(k)}\,_{(l)}({1\over r})\;,$$

\noindent such that $\,^{0,1}\omega_{(k)(l)}=\,-\,^{0,1}\omega_{(l)(k)}$,
and $\lbrace ^0\omega_{(k)(l)} \rbrace$ are constants.
Taking into account (8) it is easy to see that when 
integrated over the whole three-dimensional space the integral
on the right hand side of (35) reduces to a vanishing expression:

$${1\over {8\pi}}\int_{V\rightarrow \infty} d^3x 
(\partial_i \partial_j\, ^1\omega_{(i)(j)}
-\partial_i \partial_i\, ^1\omega_{(j)(j)})=0\;.$$

\noindent Therefore we expect to find the same result for the 
total energy if we evaluate (27) out of any element of the set of
triads (25).

\bigskip
\bigskip
\noindent {\bf VII. Discussion}\par
\bigskip
The application of the energy definition (1) for a given solution of
Einstein's equations requires considering a foliation of the 
space-time in three-dimensional spacelike surfaces.
The metric for the spacelike section of Bondi's 
radiating metric admits an infinit set of triads related by local
SO(3) transformations. In the present case, from this 
infinit set of triads we singled out two of them. We have considered
in detail the two ones that exhibit the simplest structures in
spherical coordinates, and
that: (i) satisfy
the asymptotic conditions given by (8); (ii) reduce to the 
reference space ($T_{(k)ij}=0$ everywhere) if we make the 
physical parameters vanish: $M=c=d=0$.

The two sets of triads, (20) and (22), describe the spacelike section
of Bondi's metric given by (19) and lead to the energy
expressions (28) and (32), respectively.  These expressions establish
distinct and quite definite physical predictions. They allow us to
compute the energy radiated from the interior of a spherical surface
of constant radius $r_o$.  It 
would be a remarkable achievement of the TEGR if, on physical grounds,
we could decide for one of them or even for an arbitrary element
of (25). In the TEGR 
the space (space-time) geometry is fundamentally described by 
triads  (tetrads). Unfortunately we do not dispose of experimental 
information for taking such a decision.

It is a very important result that the {\it total} energy due to
both sets of triads (as well as from any element of (25) )
agrees exactly with the static Bondi energy,
in which case the energy arises from the
integration of $M=M(\theta)$.
In fact, the definition of Bondi's 
{\it mass aspect} is basically motivated by the fact that in the
static case, and by investigating the asymptotic properties of the
gravitational field, $M(\theta)$ arises as the mass of an
isolated system. 

The final expressions
(29) and (33) support the consistency of the definition 
(1), and the relevance of the TEGR as a fundamental description of 
general relativity. However the present analysis, which was developed
on spacelike surfaces, has to be compared with the one recently
carried out on null surfaces\cite{Maluf8}. Let us recall that in order
to obtain the Hamiltonian formulation given by (5-7) we imposed
the time gauge condition. Therefore the resulting geometry may be
understood as a {\it three-dimensional} teleparallel geometry, since
the teleparallelism is restricted to the three-dimensional
spacelike surface. On the other hand, in the Hamiltonian 
formulation developed in \cite{Maluf8} we have not fixed any
particular tetrad component, and consequently the
teleparallel geometry is really {\it four-dimensional}. 

In \cite{Maluf8}
the constraints also contain a total divergence, in similarity with
(7a), and may be taken likewise to define the gravitational
energy-momentum. Although it appears that the geometrical
framework of \cite{Maluf8} is better suited to the analysis of
the Bondi-Sachs metric, we note that the the energy expression
arising there is considerably more complicated than (1). Moreover
we do not know yet whether the constraint algebra leads to a
consistent Hamiltonian formulation
(either on null or spacelike surfaces). We also note that one
Hamiltonian formulation cannot be reduced to the other by 
means of gauge fixing. Nature admits only one 
correct physical description, and therefore either the 
three-dimensional or the four-dimensional teleparallel geometry
is the correct candidate for describing the energy properties
of the gravitational field. All these issues will be
considered in the near future.   \\

\bigskip
\noindent {\it Acknowledgements}\par
\noindent We are grateful to the referee for several remarks that 
led to the improvement of the paper.
This work was supported in part by CNPQ. J.F.R.N. is
supported by CAPES, Brazil.\par

\end{document}